\documentclass[12pt]{article}   
   
\usepackage{array}   
\usepackage{epsfig}   
\usepackage{amssymb}   
\usepackage{graphics,graphpap}   
\usepackage{amssymb}   
\usepackage{amsmath}   
\usepackage[usenames]{color}  
\usepackage{slashed}   
\usepackage{graphicx}

\renewcommand{\thefootnote}{\fnsymbol{footnote}}   
   
\def\s#1{\setbox0=\hbox{$#1$}%
  \rlap{\ifdim\wd0>.7em\kern.22\wd0\else\kern.1\wd0\fi /}#1}   

   
\begin{document}   
   
\begin{titlepage}   
\begin{flushright}   
\begin{tabular}{l}   
IPPP/14/36
\\
DCPT/14/72
\end{tabular}   
\end{flushright}   
\vskip1.5cm   
\begin{center}   
{\Large \bf \boldmath Selected Topics in Heavy Flavour Physics}
\vskip1.3cm    
{\sc   
Alexander J. Lenz      \footnote{alexander.lenz@durham.ac.uk}$^{,1}$}   \vskip0.5cm   
$^1$ Institute for Particle Physics Phenomenology, Department of Physics,
     University of Durham, South Road, Durham DH1 3LE, UNITED KINGDOM
\vskip2cm

   
\vskip3cm   
   
{\large\bf Abstract\\[10pt]} \parbox[t]{\textwidth}{   
We review the status of flavour physics in spring 2014. 
The numerous accurate new measurements of flavour experiments 
have enabled us to test our theoretical understanding of 
flavour processes with an unprecedented precision. 
At first sight the dominant amount of measurements
seems to be standard model like. Having a closer look one finds, 
however, that in most of the observables there is still some 
considerable space for new effects. In addition many discrepancies
are still not settled yet.
For further investigations and definite conclusions an 
improvement of the theoretical 
precision as well as  the experimental one is mandatory.
}   
   
\vfill   
   
\end{center}   
\end{titlepage}   
   
\setcounter{footnote}{0}   
\renewcommand{\thefootnote}{\arabic{footnote}}   
\renewcommand{\theequation}{\arabic{section}.\arabic{equation}}   
   
\newpage   
   
\section{Introduction}   
\setcounter{equation}{0}   
The standard model of particle physics \cite{Glashow:1961tr,Weinberg:1967tq,Salam:1968rm}
is  finally complete. The Higgs particle that was predicted 
in 1964 \cite{Englert:1964et,Higgs:1964pj,Guralnik:1964eu}  was
found in 2012 at LHC by the ATLAS and CMS Collaborations \cite{Aad:2012tfa,Chatrchyan:2012ufa}.
Knowing the value of the mass of the Higgs particle, for the first time a complete electro-weak
precision fit could be performed without having any unmeasured standard model parameters. 
This was done in \cite{Eberhardt:2012gv} and slightly later in \cite{Baak:2012kk}
and a very good overall consistency has been found.
\\
Despite these and numerous other successes, the standard model leaves many 
open questions. Far reaching ones like the quest for the quantisation of gravity, or an 
understanding 
of dark energy. We also do not know the origin of dark matter and we might want to
answer simple sounding questions,
like, why are there three generations of matter in nature. Another very profound question is,
where does matter, i.e. an excess of matter over anti-matter in the universe come from.
Sakharov has shown already in 1967 \cite{Sakharov:1967dj} that a matter-anti-matter 
asymmetry can be
created dynamically if the fundamental laws of nature have the following basic properties:
\begin{itemize}
\item Baryon number is violated.
\item There was a  phase out-of thermal equilibrium in the early universe.
\item C and CP are violated.
\end{itemize}
Focusing on the requirement of CP-violation one finds that this effect is included in the 
standard model in the quark mixing matrix\footnote{We will not discuss here the possibility
of having CP violation also in lepton mixing or in the strong sector.}, 
the Cabibbo-Kobayashi-Maskawa matrix 
\cite{Cabibbo:1963yz,Kobayashi:1973fv}. Experimentally CP-violation was already 
found in 1964 \cite{Christenson:1964fg}
as a tiny effect in the decay of Kaons. 
Larger effects were predicted already in 1981 
\cite{Carter:1980tk,Bigi:1981qs} in the decay of $B$-mesons, in particular in the decay
$B_d \to J / \psi K_S$. 
Indirect CP violation in the $B_d$-system was then found subsequently in 2001 
by BaBar \cite{Aubert:2001nu} and Belle \cite{Abe:2001xe}.
Besides the 
$B_d$-system, where direct CP-violation was established in
2006 at Belle \cite{Ishino:2006if} in the decay $B_d \to \pi^+ \pi^-$
and in 2007 at BaBar \cite{Aubert:2007mj} in the decays $B_d \to \pi^+ \pi^-,  K^{+} \pi^{-}$,
direct CP violation has also been found by the LHCb Collaboration in 
2012 in the decay $B^+ \to DK^+ $\cite{Aaij:2012kz}
and in 2013 in the decay $B_s \to K^- \pi^+$ \cite{Aaij:2013iua}.
Indirect CP violation in the $B_s$ sector, which is predicted to be very small in the
standard model (see e.g. the review \cite{Lenz:2012mb}),
has not yet been detected, despite intense searches, e.g.
\cite{LHCb:2011aa,Aaij:2013oba}.
In 2011 there were also some indications at LHCb that there might be 
direct CP violation in the charm sector \cite{Aaij:2011in}, which were, however, 
not confirmed by more recent studies in 2013 \cite{Aaij:2013bra},
see also the discussion in \cite{Lenz:2013pwa}.
\\
One of the reasons for the interest in flavour physics stems from the fact
that meson decays are currently the only processes in nature, where CP violation 
has been detected. By studying these decays in detail, one hopes to deepen our 
understanding of the origin of CP violation. Moreover flavour physics
enables indirect searches for new physics, where very precise measurements are 
compared with very precise standard model calculations. Significant deviations 
might then point towards beyond standard model contributions. Such a programme is 
complementary to the direct searches, at e.g. the general purpose detectors 
ATLAS and CMS, where one hopes to detect decay products of directly produced 
new particles. As long as no direct evidence for new physics is found,
indirect searches might provide the first hints for new effects at a higher
energy scale than directly accessible and as soon as direct evidence for new 
particles is found, the precision study of flavour effects will be helpful
in determining the new flavour couplings.
Indirect searches rely of course heavily on our control of the corresponding 
hadronic uncertainties in flavour transitions. For the case of $b$-hadrons two
facts turn out to be very useful in that respect. First, the strong coupling at 
the scale $\mu = m_b$ is relatively small $\alpha_s(m_b) \approx 0.2$ and second,
there exists an expansion of decay rates in terms of the inverse heavy quark mass,
the heavy quark expansion (HQE) \cite{HQE}, which allows precise predictions.
Several non-trivial cross-checks for these tools to handle QCD effects 
will be discussed below. An interesting question is of course, to what extent
the HQE methods can be applied in the charm sector, where the expansion parameters 
$\alpha_s(m_c)$ and $1/m_c$ are considerably larger.
A final motivation for flavour physics studies are precise determinations of 
many standard model parameters, like the values of the CKM parameters or 
also some quark masses.
\\
After the big success of the B-factories with the detectors BaBar and Belle, see e.g.
\cite{Giorgi:2013uza}
and 
the results from TeVatron, see e.g. \cite{Borissov:2013fwa},
the field is currently dominated by the LHCb experiment
(see e.g. \cite{Bediaga:2012py} for some earlier results),
but there are also some important contribution from ATLAS, see e.g. \cite{Leontsinis:2013zza}, 
and CMS, see e.g. \cite{Mironov:2013jaa}, as will be discussed
below.
\\
In Section \ref{inclusive} we will study inclusive quantities like lifetimes, but
also the mixing system as well as individual inclusive branching ratios. Most of the
corresponding theory predictions rely on the heavy quark expansion.
In Section \ref{exclusive} we switch to exclusive quantities, starting from leptonic decays,
over semi-leptonic decays to non-leptonic ones.
In Section \ref{newphysics} we discuss some consequences for searches for new physics models
and in Section \ref{conclusion} we conclude. 

\section{Inclusive decays}
\label{inclusive}
We start our  discussion with inclusive decays. Such decays are characterised by the fact 
that we do not specify the hadronic final state, simplifying thus the non-perturbative 
physics considerably. 
The prime example of an inclusive quantity are lifetimes of $b$- and $c$-hadrons, as well
as observables related to the mixing of neutral mesons. Finally we discuss also
individual semi- and non-leptonic
inclusive decay modes.
\subsection{Lifetimes}
Lifetimes are among the most fundamental properties of a particle. We compare here recent 
measurements for the lifetime of $D$-mesons, $B$-mesons and $b$-baryons with the latest
theory predictions.
     \subsubsection{Theory}
        Total decay rates can be written according to the heavy quark expansion 
        - see \cite{HQE} for the first systematic expansion and \cite{lifetime} for a
        review of the extensive literature -  as
        \begin{equation}
         \Gamma = \Gamma_0 + \frac{\Lambda^2}{m_q^2}\Gamma_2  
                           + \frac{\Lambda^3}{m_q^3}\Gamma_3   
                           + \frac{\Lambda^4}{m_q^4}\Gamma_4 + \dots
           \; .
         \label{gammatot}
        \end{equation}
        If the mass $m_q$ of the decaying quark is heavy and the hadronic scale $\Lambda$ is 
        not very large,
        then the expansion in Eq.(\ref{gammatot}) is expected to converge quickly. In particular
        because there are no corrections of order $\Lambda/m_b$.
        Each of the coefficient $\Gamma_i$ for $i\geq2$ consists of perturbatively calculable
        Wilson-coefficients and of non-perturbative matrix-elements, that have to be determined, e.g.
        with lattice calculations or QCD sum rules.
        For more details we refer the interested reader to the review \cite{lifetime}.
     \subsubsection{Charmed mesons}
        For charmed mesons one finds experimentally a huge spread in the lifetime ratios
        \cite{PDG}
        \begin{equation}
        \frac{\tau (D^+  )}{\tau (D^0)}^{\rm Exp.}  =  2.536 \pm 0.017 \; , 
        \hspace{1cm}
        \frac{\tau (D_s^+)}{\tau (D^0)}^{\rm Exp.}  =  1.219 \pm 0.017 \; . 
        \label{tauD}
        \end{equation}
        Besides the fact that $1/m_c$ does not look like a good expansion parameter, the values
        in Eq.(\ref{tauD})  indicate huge corrections in Eq.(\ref{gammatot}), if not
        a complete breakdown of the expansion. Nevertheless, studies within the HQE
        were performed, see e.g. \cite{lifetime} for the history of these efforts.
        In \cite{Lenz:2013aua} an investigation of the $D$-meson lifetimes including 
        $\alpha_s$-corrections to $\Gamma_3$ and the LO-corrections to $\Gamma_4$ obtained
        \begin{equation}
        \frac{\tau (D^+  )}{\tau (D^0)}^{\rm HQE} =  2.2 \pm 0.4 ^{+0.3}_{-0.7}\; , 
        \hspace{1cm}
        \frac{\tau (D_s^+)}{\tau (D^0)}^{\rm HQE} =  1.19 \pm 0.12 \pm 0.04 \; . 
        \label{tauDtheory}
        \end{equation}
        The first error stems from the uncertainties in the non-perturbative matrix 
        elements of the 
        arising four-quark operators. For these matrix elements some assumptions had to be made
        in \cite{Lenz:2013aua}, since there is no first principle calculation available. Such
        an endeavour would be very desirable. The second error in Eq.(\ref{tauDtheory}) stems
        from the renormalisation scheme dependence, which could be reduced by a NNLO-QCD calculation.
        Contrary to the naive expectation the HQE seems to be capable of describing the huge 
        lifetime ratios in the $D$-meson system, but for more profound statements, lattice values
        for the arising bag parameters are mandatory.

     \subsubsection{B-mesons}
        For $B$-mesons the measured lifetime ratios are very close to one \cite{HFAG}
        \begin{equation}
        \frac{\tau (B^+)}{\tau (B_d)}^{\rm Exp.} =  1.079 \pm 0.007 \; , 
        \hspace{1cm}
        \frac{\tau (B_s)}{\tau (B_d)}^{\rm Exp.} =  0.998 \pm 0.009 \; . 
        \label{tauB}
        \end{equation}
        More recent experimental numbers, that are not yet included in the HFAG 
        average can be found in \cite{Aaij:2014owa}.
        Because of the larger value of $m_b$ one expects now a better  convergence
        of Eq.(\ref{gammatot}). Unfortunately it turns out, see e.g. the detailed discussion
        in \cite{lifetime}, that pronounced cancellations are occuring in the theory 
        predictions of these ratios, that lead to a strong sensitivity on the bag parameters
        and the most recent determination of these parameters stems already from 2001 
        \cite{Becirevic:2001fy}.
        Relying on these old non-perturbative values and including the NLO-QCD corrections
        from \cite{Beneke:2002rj,Franco:2002fc}
        one gets \cite{Lenz:2011ti}
        \begin{equation}
        \frac{\tau (B^+)}{\tau (B_d)}^{\rm HQE} =  1.044 \pm 0.024 \; , 
        \hspace{1cm}
        \frac{\tau (B_s)}{\tau (B_d)}^{\rm HQE} =  0.998 \pm 0.002 \; . 
        \label{tauBtheory}
        \end{equation}
        The ratio of the neutral mesons is in perfect agreement with data, while the prediction
        for $\tau (B^+) / \tau (B_d)$ is slightly smaller than the measurement quoted in Eq.(\ref{tauB}),
        but for more far-reaching statements precise bag parameters are urgently needed.
        Since the lifetime ratio $\tau (B_s)/ \tau (B_d)$ is affected by very pronounced cancellations
        in the standard model, this quantity can also be used as an important bound on 
        hidden $B$-decay channels due to new physics, see e.g. the recent investigation in
        \cite{Bobeth:2014rda}.
     \subsubsection{$b$-baryons}
        The $\Lambda_b$ lifetime suffered from a longstanding discrepancy
        between experiment and theory that was finally settled experimentally.
        HFAG gave in 2003 an average of 
        \begin{equation}
        \frac{\tau (\Lambda_d)}{\tau (B_d)}^{\rm HFAG \, 2003} =  0.80  \pm 0.05 \; .
        \label{tauLambdab2003}
        \end{equation}
        Older numbers resulted in even smaller ratios. The value in Eq.(\ref{tauLambdab2003})
        was in disagreement
        with early estimates based on the HQE, see e.g. \cite{Shifman:1986mx}
        (see \cite{lifetime} for a more
        detailed history of prediction of the $\Lambda_b$ lifetime)
        \begin{equation}
        \frac{\tau (\Lambda_d)}{\tau (B_d)}^{\rm HQE \, 1986} \approx  0.96 \; .
        \label{tauLambdabtheoryold}
        \end{equation}
        Again the theory prediction depends strongly on the value of the non-perturbative
        matrix elements and in this case we have only an exploratory lattice study from
        1999 \cite{DiPierro:1999tb}, which yielded quite large numerical values
        for the bag parameters, leading to a larger deviation of the ratio from one.
        Taking these numbers and also looking for some additional effects that might reduce
        the ratio, one could arrive at values as low as \cite{Gabbiani:2004tp}
        \begin{equation}
        \frac{\tau (\Lambda_d)}{\tau (B_d)}^{\rm HQE \, 2004} =  0.86  \pm 0.05 \; .
        \label{tauLambdabtheorymiddle}
        \end{equation}
        In recent years there were a lot of new measurements from 
        CDF \cite{Aaltonen:2014wfa}
        and D0 \cite{Abazov:2012iy}
        at TeVatron  and 
        also from CMS \cite{Chatrchyan:2013sxa}, 
        ATLAS \cite{Aad:2012bpa}
        and of course
        LHCb \cite{Aaij:2014owa,Aaij:2014zyy}
        that found considerably higher values for the $\Lambda_b$-lifetime.
        The current HFAG \cite{HFAG} average reads
        \begin{equation}
        \frac{\tau (\Lambda_d)}{\tau (B_d)}^{\rm HFAG \, 2013} =  0.941  \pm 0.016 \; .
        \label{tauLambdab2013}
        \end{equation}
        In \cite{lifetime} the $\Lambda_b$ lifetime was re-investigated, using 
        spectroscopic information for matrix elements (following  \cite{Rosner:1996fy})
        and the NLO-QCD result from \cite{Franco:2002fc}, as well as the $1/m_b$ corrections
        from \cite{Gabbiani:2004tp}, with the result
        \begin{equation}
        \frac{\tau (\Lambda_d)}{\tau (B_d)}^{\rm HQE \, 2014} =  0.95  \pm 0.05 \pm ??? \; .
        \label{tauLambdabtheorynew}
        \end{equation}
        The final number depends, however, crucially on the precise value of the bag parameter,
        where we are lacking a first principle calculation, thus we added the question marks.
     \subsubsection{Lifetime upshot}
       The above comparison between experiment and theory shows that the HQE seems to work 
       well for lifetimes of heavy hadrons, even in the case of $D$-mesons. For more precise 
       statements new lattice investigations are urgently needed.
       Further examples like the $B_c$ meson lifetime and the $\Xi_b$ lifetime are discussed 
       in the review \cite{lifetime}.
  \subsection{Mixing Quantities}
\begin{figure}[h]
  \centering
  \fbox{
    \includegraphics[width=0.9\textwidth]{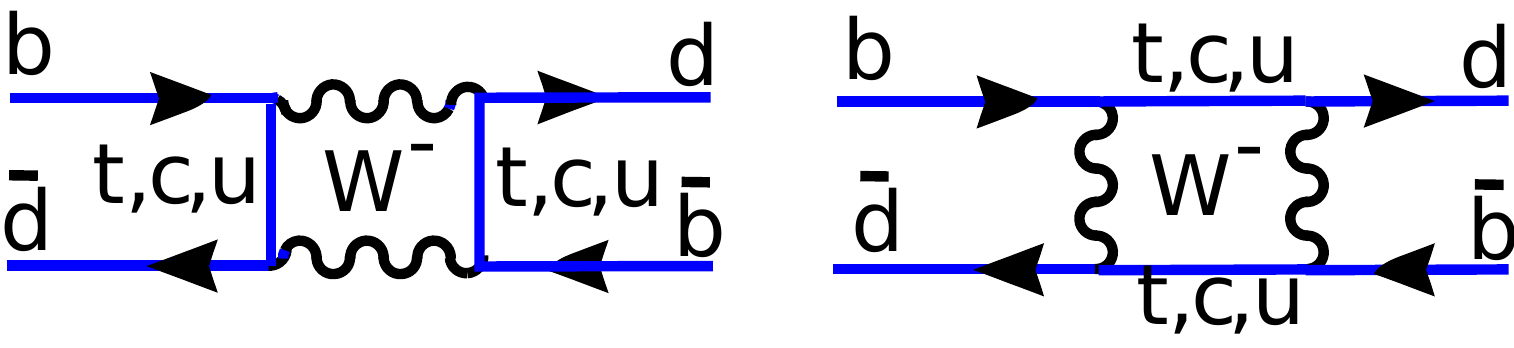}
  }
  \caption{Box diagrams contributing to the mixing of $B_d$-mesons. The diagrams on the l.h.s. 
can only contribute to $M_{12}^q$, because the $W$-bosons are always off-shell, this is also 
the case for the top-quark contribution of the diagram on the r.h.s. . 
Contributions to $\Gamma_{12}^q$ can only arise from the up- and charm-quark on the r.h.s. .}
\label{fig:box}
\end{figure}

The phenomenon of particle-antiparticle mixing is a macroscopic quantum effect.
It arises due to so-called box diagrams
shown in Fig. \ref{fig:box}, which enable a transition of a neutral meson state, defined by its
quark flavour content into its anti-particle.
This effect shows that the flavour eigenstates of the neutral mesons, e.g. $B_d = (\bar{b}d)$
and $\bar{B}_d = (b \bar{d})$, do not coincide with the mass eigenstates, which we denote
by $B_H$ and $B_L$, where $H$ stands for heavy and $L$ for light. In the Kaon system the notation
$K_S$ and $K_L$ is used, where $S$ stands for short-lived and $L$ for long-lived.
Performing a change of basis, one finds a mass difference $\Delta M_d$ and a decay rate
difference $\Delta \Gamma_d$ of the mass eigenstates.
\begin{eqnarray}
\Delta M_d & := & M_H - M_L  \; ,
\\
\Delta \Gamma_d & := & \Gamma_L - \Gamma_H  \; ,
\end{eqnarray}
where $M_H$ denotes the mass of the heavy eigenstate a.s.o..
\subsubsection{A very brief history of mixing}
Mixing is by now well established in several systems of neutral mesons:
\begin{itemize}
\item[1956] $K^0$-system: Mixing in the neutral $K$-system was theoretically developed in 
            1955 by Gell-Mann and Pais \cite{GellMann:1955jx}. Based on that framework
            the phenomenon of {\it regeneration} was predicted in the same year by Pais and 
            Piccioni  \cite{Pais:1955sm}. Experimentally regeneration was confirmed in 
            1960 \cite{Muller:1960ph}. A huge lifetime difference between the two neutral
            $K$-mesons was established already in 1956 \cite{Lande:1956pf}.
\item[1986] $B_d$-system: Mixing in the $B_d$-system was found 1986 by UA1 at 
            CERN \cite{Albajar:1986it} 
            (UA1 attributed the result however to $B_s$ mixing)
            and 1987 by ARGUS at DESY \cite{Albrecht:1987dr}. The large result for
            the mass difference $\Delta M_d$ can be seen as the first clear hint for an 
            (at that time) unexpected large value of the top quark mass \cite{Ellis:1987mm}
            \footnote{To avoid a very large value of the top quark mass, also different new physics
            scenarios were investigated, in particular a scenario with a heavy fourth generation
            of fermions and a top quark mass of the order of 50 GeV, see e.g. 
            \cite{Tanimoto:1987wv}.}.
            For the decay rate difference - which is expected to have a small value in the 
            standard model \cite{Lenz:2011ti,Bobeth:2014rda} - currently only upper bounds 
            are available from  BaBar \cite{Aubert:2006nf}, Belle \cite{Higuchi:2012kx}, 
            D0 \cite{Abazov:2013uma} and LHCb \cite{Aaij:2014owa}. 
            Here further experimental studies are very welcome, because in this quantity there 
            is still room for some sizable new physics effects \cite{Bobeth:2014rda}.
\item[2006/12] $B_s$-system: The large mass difference in the $B_s$-system was established
            in 2006 by the CDF
            Collaboration at TeVatron \cite{Abulencia:2006ze}. In 2012 the LHCb Collaboration
            measured for the first time a non-vanishing value of the decay rate difference in the 
            $B_s$-system \cite{LHCb:2012}.
\item[2007] $D^0$-system: Here we had several experimental evidences 
            (BaBar, Belle, Cleo, CDF, E791, E831 FOCUS, LHCb)
            for values of
            $\Delta \Gamma / \Gamma$ and $\Delta M / \Gamma$ at the per cent level,
            but the first single measurement with a statistical significance
            of more than five standard deviations was done only in 2012
            by the LHCb collaboration \cite{Aaij:2012nva}.

\end{itemize}

     \subsubsection{Theory}
Mass differences and decay rate differences of neutral $B$-mesons
can be expressed to a very high accuracy as (see e.g. \cite{Lenz:2012mb}
for the explicit form of the tiny corrections)
\begin{eqnarray}
\Delta M_q & = & 2 | M_{12}^q | \; ,
\\
\Delta \Gamma_q & = & 2 |\Gamma_{12}^q| \cos (\phi_q) \; ,
\label{deltagamma}
\end{eqnarray}
with $M_{12}^q$ being the dispersive part of the box diagrams, 
see Fig.\ref{fig:box},
and $\Gamma_{12}^q$ being the absorptive part. The mixing phase reads
$\phi_q = \arg\left(- M_{12}^q/ \Gamma_{12}^q \right)$.
The dispersive part $M_{12}^q$ is sensitive to off-shell intermediate states; 
in the case of the neutral $B$ mesons, the by far largest contribution stems from the 
virtual top quark in the loop. This part is also very sensitive to hypothetical 
heavy new physics particles in the loop.
For the $B$-meson system one gets after integrating out the heavy $W$-boson and the top-quark
       \begin{eqnarray}
        M_{12}^q & = & \frac{G_F^2}{12 \pi^2} 
          (V_{tq}^* V_{tb})^2 M_W^2 S_0(x_t)
          { B_{B_q} f_{B_q}^2  M_{B_q}} \hat{\eta }_B \; .
        \label{M12SM}
        \end{eqnarray}
The evaluation of the 1-loop box-diagram gives the so-called Inami-Lim function $S_0(x_t)$
\cite{Inami:1980fz} with $x_t = m_t^2/M_W^2$.
Perturbative QCD corrections to the box-diagrams are denoted by { $\hat{\eta }_B$}
\cite{Buras:1990fn,Urban:1997gw}, they turned out to be  ample.
The arising non-perturbative matrix element of the four quark operator $Q$ 
is for historical reasons parameterised in terms of a bag parameter $B_{B_q}$ and a decay constant
$f_{B_q}$:
\begin{eqnarray}
\langle \bar{B_q}|
Q
|B_q \rangle
& = & 
\frac{8}{3} { B_{B_q} f_{B_q}^2  M_{B_q}^2} \; ,
\label{Bag}
\end{eqnarray}
with the operator
\begin{eqnarray}
Q & = & \bar{q}_\alpha \gamma_\mu (1- \gamma_5) b_\alpha
\cdot
\bar{q}_\beta \gamma^\mu (1-\gamma_5) b_\beta \; .
\label{Q}
\end{eqnarray}
$\alpha$ and $\beta$ denote colour indices.
The bag parameter and the decay constant have to be determined with non-perturbative methods like
lattice QCD or QCD sum rules.
\\
$\Gamma_{12}^q$ is sensitive to on-shell intermediate states; thus only
the up- and charm-quark on the r.h.s. of Fig.\ref{fig:box} can contribute.
After integrating out the heavy $W$-bosons one performs a second 
operator-product expansion (OPE), 
the HQE, yielding a similar form as in Eq.(\ref{gammatot})
\begin{equation} 
\Gamma_{12} = \frac{\Lambda^3}{m_b^3} \left( \Gamma_3^{(0)} + 
              \frac{\alpha_s(\mu)}{4 \pi} \Gamma_3^{(1)} + \dots \right) 
              + \frac{\Lambda^4}{m_b^4} \left( \Gamma_4^{(0)} + \dots \right) + \dots \; .
\end{equation}
Again $\Gamma_i$ consists of perturbative Wilson coefficients and non-perturbative matrix elements.
Because of the arising CKM matrix elements both $M_{12}^q$ and $\Gamma_{12}^q$ 
can be complex.
     \subsubsection{Charm mixing}
The mixing observables in the charm system are typically denoted as $x$ and $y$
\begin{equation}
x = \frac{\Delta M}{\Gamma} \; , \hspace{1cm}
y = \frac{\Delta \Gamma}{2 \Gamma} \; .
\end{equation}
In contrast to $B$-mixing, where we have the very heavy top-quark, as well as
the charm- and the up-quark as virtual loop particles, charm mixing proceeds via 
internal $d$-,$s$- and $b$-quarks.
The lower masses of the internal and external particles lead to the fact that
the methods used for the determination of the mixing observables in the $B$-system
are much less justified for describing $D$-oscillations.
\\
Because of the promising result of the investigations of $D$-meson lifetimes
one might try nevertheless to use the HQE for a determination of $\Gamma_{12}$,
as it was done in \cite{Bobrowski:2010xg}. In that case, however, a second problem
arises. The leading term in the HQE, $\Gamma_3$ suffers from an almost perfect GIM
\cite{Glashow:1970gm} cancellation.
Thus the idea came up quite some time ago \cite{Georgi:1992as,Ohl:1992sr} that $D$-mixing is 
described by higher orders in the HQE, i.e. $\Gamma_6$ and $\Gamma_9$, where
the GIM cancellation is much less pronounced, see \cite{Bigi:2000wn}. Until now no 
satisfactionary calculation of these higher order effects was performed.
The conclusion of \cite{Bigi:2000wn} was that standard contributions to $x$ and $y$ 
of up to $1 \%$ are not excluded, while \cite{Bobrowski:2010xg} concluded that they are 
probably smaller. It was also shown in \cite{Bobrowski:2010xg} that the enhancement effect
suggested in \cite{Georgi:1992as,Ohl:1992sr,Bigi:2000wn} could also lead to CP violating
effects
in mixing of the order of several per mille. But here clearly more work has to be done.
Because of these drawbacks it was also tried to use an exclusive approach in order
to describe charm mixing \cite{Falk:2001hx,Falk:2004wg}, leading to a similar conclusion:
$x$ and $y$ might have values of about $1\%$ in the standard model.
\\
Experimentally $D$-mixing is now well settled. HFAG \cite{HFAG} quotes as averages
\begin{equation}
x^{\rm Exp.} = \left(0.39^{+0.16}_{-0.17}  \right) \%
\; , \hspace{1cm}
y^{\rm Exp.} = \left( 0.67^{+0.07}_{-0.08} \right)  \%
\; ,
\end{equation}
while CP violation in $D$-mixing is still quite weakly constrained \cite{Aaij:2013ria}, 
which will, however, change in future, see e.g. \cite{Contu:2013rna}. 
\\
Despite the drawbacks related to our insufficient understanding of the standard model
contribution, the $D$-mixing system is, however very well suited to look for new 
physics effects, because the contribution of heavy new particles, can be 
calculated more reliably, see e.g. \cite{Golowich:2007ka,Gedalia:2009kh}.

\subsubsection{B-mixing}
       
In $B$-mixing the theory is under much better control and we predict for the mass differences
\cite{Lenz:2011ti}
\begin{equation}
\Delta M_d^{\rm Theory} = \left( 0.543 \pm 0.091 \right)\, {\rm ps}^{-1} \, ,
\hspace{1cm}
\Delta M_s^{\rm Theory} = \left( 17.3 \pm 2.6 \, \right){\rm ps}^{-1} \, .
\end{equation}
The large theory uncertainty is dominated by the values of the hadronic matrix elements. We have
used the most recent result from FLAG \cite{Colangelo:2010et}
for $f_{B_q} B^2$, which is simply the result from \cite{Gamiz:2009ku}. Similar values were obtained
in \cite{Carrasco:2013zta} and slightly higher ones in \cite{Bouchard:2011xj}.
These predictions can be compared with the most recent experimental averages from 
HFAG \cite{HFAG}\footnote{The
most precise measurements for 
$\Delta M_d$ \cite{Aaij:2012nt}
and
$\Delta M_s$ \cite{Aaij:2013mpa}
are currently obtained from the LHCb collaboration.}
\begin{equation}
\Delta M_d^{\rm Exp.} = \left(0.510 \pm 0.004 \right)\, {\rm ps}^{-1} \, ,
\hspace{1cm}
\Delta M_s^{\rm Exp.} = \left(17.69 \pm 0.08 \right) \, {\rm ps}^{-1} \, .
\end{equation}
The central values agree perfectly with the standard model predictions, but due to the large
theory uncertainties there is still some room for new physics effects.
\\
The calculation of the decay rate difference relies on the HQE, which was questioned
in particular for the case of $\Delta \Gamma_s$, which is governed by the quark level decay 
$ b \to c \bar{c} s$. In that case the energy release is substantially limited compared to
a $b$-decay into mass-less final states and thus the expansion parameter of the HQE naively seems
to be large.
An explicit calculation including
NLO-QCD corrections \cite{Beneke:1998sy,Beneke:2003az,Ciuchini:2003ww,Lenz:2006hd}
and
subleading HQE corrections \cite{Beneke:1996gn,Dighe:2001gc}
gives \cite{Bobeth:2014rda,Lenz:2011ti}
\begin{equation}
\Delta \Gamma_d^{\rm HQE} = \left(0.0029 \pm 0.0007 \right) \, {\rm ps}^{-1} \, ,
\hspace{1cm}
\Delta \Gamma_s^{\rm HQE} = \left(0.087 \pm 0.021 \right)\, {\rm ps}^{-1} \, .
\end{equation}
$\Delta \Gamma_s$ was measured for the first time in 2012 by the LHCb
Collaboration \cite{LHCb:2012}. The current average from HFAG \cite{HFAG}
reads
\begin{equation}
\Delta \Gamma_s^{\rm Exp.} = \left( 0.081 \pm 0.011 \right)\, {\rm ps}^{-1} \, ,
\end{equation}
it includes the measurements from LHCb \cite{Aaij:2013oba}, ATLAS \cite{Aad:2012kba},
CDF \cite{Aaltonen:2012ie} and D0 \cite{Abazov:2011ry}.
Experiment and theory agree perfectly for $\Delta \Gamma_s$, excluding thus huge 
violations of quark hadron duality. The experimental uncertainty will be reduced in future,
while the larger theory uncertainty is dominated from unknown matrix elements of 
dimension seven operators, see \cite{Lenz:2006hd,Lenz:2011ti}. Here a first lattice 
investigation or a continuation of the QCD sum rule study in \cite{Mannel:2007am,Mannel:2011zza}
would be very welcome.
\\
$\Delta \Gamma_d$ has not been measured yet. The HFAG average \cite{HFAG} includes measurements
from  BaBar \cite{Aubert:2006nf} and Belle \cite{Higuchi:2012kx}, but there were also
two investigations from D0 \cite{Abazov:2013uma} and LHCb \cite{Aaij:2014owa}.
LHCb compared the difference in the effective lifetimes of $B_d \to J / \psi K^*$ and
 $B_d \to J / \psi K_S$, while D0 found that $\Delta \Gamma_d$ can give a sizable
contribution 
\cite{Borissov:2013wwa}
to the dimuon asymmetry
\cite{Abazov:2010hv,Abazov:2010hj,Abazov:2011yk,Abazov:2013uma}. 
The different values read 
\begin{eqnarray}
\frac{\Delta \Gamma_d}{\Gamma_d}^{\rm HFAG} &= &  \left(1.5  \pm 1.8 \right)\% \, ,
\\
\frac{\Delta \Gamma_d}{\Gamma_d}^{\rm D0}   &= &  \left(0.50 \pm 1.38 \right)\%  \, ,
\\
\frac{\Delta \Gamma_d}{\Gamma_d}^{\rm LHCb} &= & \left(-4.4  \pm 2.7 \right)\% \, .
\end{eqnarray}
All these bounds are compatible with the small standard model prediction
\cite{Lenz:2011ti}
\begin{eqnarray}
\frac{\Delta \Gamma_d}{\Gamma_d}^{\rm HQE} &= &  \left(0.42  \pm 0.08 \right)\% \, ,
\end{eqnarray}
but they also leave a lot of space for beyond standard model effects.
It is interesting to note that the long-standing problem of the dimuon asymmetry
\cite{Abazov:2010hv,Abazov:2010hj,Abazov:2011yk,Abazov:2013uma} could be solved by
a large value of the $\Delta \Gamma_d$, i.e.  $\Delta \Gamma_d = (6.3 \pm 1.6) \cdot
\Delta \Gamma_d^{\rm SM}$. In a model-independent study \cite{Bobeth:2014rda}
it was shown that large enhancements of $\Delta \Gamma_d$ do not violate any
other experimental bounds, which is in contrast to the situation with $\Delta \Gamma_s$, 
where no enhancement being considerably larger than the hadronic uncertainties is possible,
see e.g. \cite{Bobeth:2011st}. Therefore it would be very eligible to have more precise
experimental bounds on this quantity.
\\
There is also a third class of observables in the mixing systems, related to CP violation, 
the so-called flavour-specific or semi-leptonic asymmetries. They are defined as
\begin{eqnarray}
a_{sl}^q & = & \frac{\Gamma (\bar{B}_q(t) \to f) - \Gamma ({B}_q(t) \to \bar{f}) }
                    {\Gamma (\bar{B}_q(t) \to f) + \Gamma ({B}_q(t) \to \bar{f}) }
          =  \left| \frac{\Gamma_{12}^q}{M_{12}^q} \right| \sin \phi_q \; ,
\end{eqnarray}
where $f$ denotes a flavour-specific final state - semi-leptonic states are a special case
of flavour specific ones. 
In the standard model these asymmetries are tiny \cite{Lenz:2011ti}
\begin{equation}
a_{sl}^{d, \rm HQE} =    (-4.1 \pm 0.6) \cdot 10^{-4} \; ,
\hspace{1cm}
a_{sl}^{s, \rm HQE} =    (+1.9 \pm 0.3) \cdot 10^{-5} \; .
\end{equation}
The first measurements of the dimuon asymmetry \cite{Abazov:2010hv,Abazov:2010hj,Abazov:2011yk}
pointed towards a large enhancement of the semi-leptonic asymmetries. At that time
the measured asymmetry $A_{CP}$ was interpreted as having only contributions from CP
violation in mixing:
\begin{equation}
A_{CP} \propto A_{sl}^b = C_d a_{sl}^d + C_s a_{sl}^s \; .
\end{equation}
The first measurement of $A_{CP}$ \cite{Abazov:2010hv,Abazov:2010hj} 
was a factor of 42 larger\footnote{See
e.g. \cite{42} for the profound implications of this enhancement factor.}
than the standard model prediction in  \cite{Lenz:2006hd}. A successive
measurement \cite{Abazov:2011yk} gave a slightly smaller value, but the statistical significance 
of the deviation increased to 3.9 standard deviations. The findings from the D0
Collaboration can be tested by individual measurements of the semi-leptonic asymmetries,
which have been performed
for the $B_d$-system by D0 \cite{Abazov:2012hha} and BaBar \cite{Lees:2013sua}
and 
for the $B_s$-system by D0 \cite{Abazov:2012zz} and LHCb \cite{Aaij:2013gta}.
\begin{eqnarray}
a_{sl}^{d, \rm D0}    =  \left(+0.68 \pm 0.45 \pm 0.14   \right)     \% \; ,
&&
a_{sl}^{s, \rm D0}    =  \left(-1.12 \pm 0.74 \pm 0.17   \right)     \% \; ,
\\
a_{sl}^{d, \rm BaBar} =  \left(+0.06 \pm 0.17^{+0.38}_{-0.23} \right)\% \; ,
&&
a_{sl}^{s, \rm LHCb}  =  \left(-0.06 \pm 0.50 \pm 0.36      \right)  \% \; .
\end{eqnarray}
These numbers are consistent with the standard model predictions, but because of the
still sizable uncertainties they also do not exclude the large enhancement
of the dimuon asymmetry. 
Last year Borissov and Hoeneisen \cite{Borissov:2013wwa}
identified a new source contributing to the
measured value of $A_{CP}$, leading to 
\begin{equation}
A_{CP} \propto A_{sl}^b + C_{\Gamma_d} \frac{\Delta \Gamma_d}{\Gamma_d} 
                        + C_{\Gamma_s} \frac{\Delta \Gamma_s}{\Gamma_s} 
 \; .
\label{Master}
\end{equation}
The contribution due to $\Delta \Gamma_s$ turns out to be negligible, but
even the tiny standard model value of $\Delta \Gamma_d^{\rm SM}$ 
gives a sizable share.
Investigating different regions for the muon impact parameter separately,
it is possible to extract individual values for
$a_{sl}^d$, $a_{sl}^s$ and $\Delta \Gamma_d$ from the D0 measurements 
\cite{Abazov:2013uma}:
\begin{align}
  a_{sl}^{d, \rm D0} & = (-0.62 \pm 0.43) \% \,, &
  a_{sl}^{s, \rm D0} & = (-0.82 \pm 0.99) \% \,, &
  \frac{\Delta \Gamma_d}{\Gamma_d}^{\rm D0} & = (0.50 \pm 1.38) \% \,.
\end{align}
This result differs from the combined standard model expectation for the three 
observables by $3.0\, \sigma$. If one instead assumes that the 
semi-leptonic asymmetries $a_{sl}^d$ and $a_{sl}^s$ are given by their standard model 
values, then the decay rate difference $\Delta \Gamma_d$ measured by \cite{Abazov:2013uma} using
(\ref{Master}) is
\begin{align} \label{eq:D0new}
  \frac{\Delta \Gamma_d}{\Gamma_d}^{\rm D0} & = (2.63 \pm 0.66) \% \; ,
      \end{align}
which differs by $3.3 \, \sigma$ from the SM prediction.

\subsubsection{Mixing upshot}
The mixing observables $\Delta M_d$, $\Delta M_s$ and $\Delta \Gamma_s$ in the $B$-system
agree well with the standard model predictions. Because of some sizable
hadronic uncertainties there is still plenty of room for new physics effects.
In the case of the semi-leptonic asymmetries the current 
experimental values are compatible with the tiny standard model expectations, but
the uncertainties of the measurements are still one ($a_{sl}^d$) to two orders  
($a_{sl}^s$) of magnitude larger than the central values of the standard model predictions. 
The longstanding 
discrepancy in the dimuon asymmetry might point towards some new physics effects in
$a_{sl}^d$, $a_{sl}^s$  and $\Delta \Gamma_d$, a possibility that is currently not 
excluded by any other experimental constraint.
\\
In the charm system similar statements cannot be made because of the largely unknown size
of the standard model contribution. Here an improvement in our theoretical
understanding is very desirable. As long as this does not happen, it is not excluded
that new physics was already found in $D$-mixing and we simply could not identify it.

\subsection{Inclusive decays}
Inclusive quark decays rely on the same theoretical footing as the lifetimes, the HQE, 
which seems to be well tested now. These decays are experimentally difficult to study, 
but they might 
enable searches for hidden decay channels of heavy hadrons, see e.g. \cite{Krinner:2013cja}.
     \subsubsection{Theory}
NLO-QCD corrections turned out to be crucial for the inclusive
$b$-quark decays, see e.g. \cite{Voloshin:1994sn}. They were determined 
for $b \to c l^- \bar{\nu}$ already in 1983 \cite{Hokim:1983yt},
for $b \to c \bar{u} d   $ in 1994 \cite{Bagan:1994zd},
for $b \to c \bar{c} s   $ in 1995 \cite{Bagan:1995yf},
for $b \to $ no charm      in 1997 \cite{Lenz:1997aa}
and 
for $b \to s g $           in 2000 \cite{Greub:2000sy,Greub:2000an}.
Since there were several misprints in \cite{Bagan:1995yf} - leading to IR divergent expressions -, 
the corresponding calculation was redone in \cite{Krinner:2013cja} and the numerical result was 
updated.\footnote{The authors of \cite{Bagan:1995yf} left particle physics and it was not possible
to obtain the correct analytic expressions. The numerical results in \cite{Bagan:1995yf} were, 
however, correct.}
NNLO corrections for semi-leptonic decays have been calculated in
\cite{vanRitbergen:1999gs,Melnikov:2008qs,Pak:2008qt,Pak:2008cp,Bonciani:2008wf,Biswas:2009rb}
and some first investigations for non-leptonic decays were done in
\cite{Czarnecki:2005vr}. A complete NNLO-QCD study of the non-leptonic decays seems to
be doable now, also because the $\Delta B=1$-Wilson coefficients are known at 
NNLO precision \cite{Gorbahn:2004my}.
     \subsubsection{Semi-leptonic and radiative decays}
       Inclusive semi-leptonic decays can be used for the determination of the
       CKM elements $ V_{cb}$ and $V_{ub}$, see the PDG \cite{PDG} article
       {\it Semi-leptonic B meson decays and the determination of $V_{cb}$ and $V_{ub}$}.
       The current values for these CKM elements read \cite{PDG}
       \begin{eqnarray} 
       V_{cb}^{\rm Inclusive} = (42.4 \pm 0.9) \cdot 10^{-3}
       \; ,
       \hspace{1cm}
       V_{ub}^{\rm Inclusive} = (4.41 \pm 0.15^{+0.15}_{-0.17}) \cdot 10^{-3}
       \; .
       \end{eqnarray}
       These values are larger than the values obtained by investigating exclusive
       decays like $\bar{B} \to D^* l \bar{\nu}_l$ or $\bar{B} \to \pi l \bar{\nu}_l$ 
       where one gets \cite{PDG} 
       \begin{eqnarray} 
       V_{cb}^{\rm Exclusive} = (39.5 \pm 0.8) \cdot 10^{-3}
       \; ,
       \hspace{1cm}
       V_{ub}^{\rm Exclusive} = (3.23 \pm 0.31) \cdot 10^{-3}
       \; .
       \end{eqnarray}
       Currently it is not clear what the origin of this longstanding discrepancy is, see
       e.g. \cite{PDG} for the discussion of experimental issues and problems related 
       to estimating the hadronic uncertainties, but also for some ideas how new physics
       could be responsible for the shift.
       A more recent experimental investigation at Belle \cite{Sibidanov:2013rkk}  in 2013
       yielded the result 
       \begin{eqnarray} 
       V_{ub}^{\rm Exclusive} = (3.52 \pm 0.29) \cdot 10^{-3}
       \; .
       \end{eqnarray}
      \\
       A further related observable is the semi-leptonic branching ratio. Its standard
       model value reads \cite{Krinner:2013cja} 
       \begin{equation}
        Br_{sl}^{\rm HQE} = \frac{\Gamma ( b \to c e^- \bar{\nu}_e)^{\rm HQE}}{\Gamma_{tot}^{\rm HQE}} =
        \left( 11.6 \pm 0.8 \right) \% \; ,
       \end{equation}
       which can be compared to the following experimental values 
       \begin{eqnarray}
        Br_{sl} (B_d)^{\rm Exp.}&= &\left( 10.33 \pm 0.28 \right) \% \; ,
        \\
        Br_{sl} (B^+)^{\rm Exp.}&= &\left( 10.99 \pm 0.28 \right) \% \; ,
        \\
        Br_{sl} (B_s)^{\rm Exp.}&= &\left( 10.61 \pm 0.89 \right) \% \; ,
       \end{eqnarray}
       where the first two values are taken from the PDG \cite{PDG} and the value for
       $B_s$ is from \cite{Oswald:2012yx}. These numbers agree well with the theory
       prediction, which will be probably affected notably by the inclusion of NNLO-QCD effects.
       \\
       Finally we would like to mention the penguin induced decay $ b \to s \gamma$, that is 
       quite well measured \cite{HFAG}
       \begin{equation}
       Br(b \to s \gamma)^{\rm HFAG \, 2013} = (3.43 \pm 0.21 \pm 0.07) \cdot 10^{-4}
       \end{equation}
       and agrees well with the standard model prediction of \cite{Misiak:2006zs}
       \begin{equation}
       Br(b \to s \gamma)^{\rm Theory} = (3.15 \pm 0.23) \cdot 10^{-4} \; .
       \end{equation}
       This decay gives serious constraints on different extensions of the standard model, like 
       Two-Higgs-Doublet models or Supersymmetry.
     \subsubsection{Non-leptonic decays}
       Non-leptonic inclusive decays are not well studied experimentally, they might, however,
       be interesting for searching for new effects in a model and even decay channel
       independent way, see e.g. \cite{Krinner:2013cja}. The updated theory predictions
       read for $b \to c$ transitions
       \begin{eqnarray}
       Br(b \to c \bar{u} d)^{\rm HQE}	       & = & 0.446  \pm	0.014 \; ,
       \\
       Br(b \to c \bar{c}s)^{\rm HQE}            & = & 0.232  \pm	0.007 \; ,
       \\
       Br(b \to c e \bar{\nu}_e)^{\rm HQE}       & = & 0.116  \pm  0.008 \; ,
       \\
       Br(b \to c \mu \bar{\nu}_\mu)^{\rm HQE}   & = & 0.116  \pm  0.008 \; ,
       \\
       Br(b \to c \tau \bar{\nu}_\tau)^{\rm HQE} & = & 0.027  \pm  0.001 \; ,
       \\
       Br(b \to c \bar{u}s)^{\rm HQE}            & = & 0.024  \pm	0.001 \; ,
       \\
       Br(b \to c \bar{c}d )^{\rm HQE}           & = & 0.0126 \pm	0.0005 \; ,
       \\
       Br(b \to u \bar{u}d)^{\rm HQE}            & = & 0.0063 \pm 0.0018 \; 
       \end{eqnarray}
       and for subleading $b \to u$-transitions or penguins
       \begin{eqnarray}
       Br(b \to sg)^{\rm HQE}		       & = & 0.0050 \pm	0.0009 \; ,
       \\
       Br(b \to u \bar{c}s)^{\rm HQE}	       & = & 0.0043 \pm	0.0012 \; ,
       \\
       Br(b \to u \bar{u}s)^{\rm HQE}	       & = & 0.0024 \pm	0.0012 \; ,
       \\
       Br(b \to d \bar{d}s)^{\rm HQE}	       & = & 0.0022 \pm 0.0011 \; ,
       \\
       Br(b \to s \bar{s}s)^{\rm HQE}	       & = & 0.0018 \pm 0.0009 \; ,
       \\
       Br(b \to u e\bar{\nu}_e)^{\rm HQE}       & = & 0.0017 \pm	0.0005 \; ,
       \\ 
       Br(b \to u \mu \bar{\nu}_\mu)^{\rm HQE}   & = & 0.0017 \pm	0.0005 \; ,
       \\
       Br(b \to u \tau\bar{\nu}_\tau)^{\rm HQE}  & = & 0.0006 \pm	0.0002 \; ,
       \\
       Br(b \to dg)^{\rm HQE}		       & = & 0.00024\pm 0.00010 \; ,
       \\
       Br(b \to u \bar{c}d)^{\rm HQE}	       & = & 0.00023\pm	0.00007 \; ,
       \\
       Br(b \to s \bar{s}d)^{\rm HQE}	       & = & 0.00009\pm 0.00006 \; ,
       \\
       Br(b \to d \bar{d}d)^{\rm HQE}	       & = & 0.00008\pm	0.00005 \; .
       \end{eqnarray}
	
     \subsubsection{Inclusive upshot}
       The theory of inclusive decays is theoretically quite solid. There is, however,
       the longstanding discrepancy in the extraction of the CKM elements $V_{ub}$ and 
       $V_{cb}$, which has to be settled by further experimental and theoretical 
       investigations.
       Non-leptonic inclusive decays might provide a complementary testing 
       ground for beyond standard model
       effects. Here any experimental investigation would be very welcome. On the theory side
       the extension to NNLO-QCD seems to be worthwhile and doable.
\section{Exclusive Decays}
\label{exclusive}
We present here certain exclusive decays, that seem to be very promising in searching for new physics
effects or determining standard model parameters. We start with leptonic decays, that have the simplest
hadronic structure, because they only depend on a decay constant. Next we discuss semi-leptonic decays 
that depend  on form factors and finally we briefly discuss non-leptonic decays, where some additional
assumptions, like QCD factorisation \cite{Beneke:1999br,Beneke:2000ry,Beneke:2001ev}, 
have to be made in order to describe them theoretically.
  \subsection{Leptonic decays}
The decay $ B \to \tau \nu$ proceeds in the standard model via an annihilation into a $W$-boson.
If there exists, e.g. an extended Higgs sector, the $W$-boson could simply be replaced by a 
charged Higgs-boson.
For quite some time the experimental value of the corresponding branching ratio was about three
standard deviations above the theory prediction \cite{CKMfitter} 
(see also \cite{UTfit} for similar results)
of
\begin{equation}
Br( B^+ \to \tau^+ \nu_\tau)^{\rm SM} = (0.739^{+0.091}_{-0.071}) \cdot 10^{-4} \; .
\end{equation}
A new measurement at Belle \cite{Adachi:2012mm} found, using a hadronic tagging method
          \begin{equation}
Br( B^+ \to \tau^+ \nu_\tau)^{\rm Belle} = (0.72^{+0.29}_{-0.27}) \cdot 10^{-4} \; ,
\end{equation}
which is now perfectly consistent with the standard model expectation. An independent
confirmation of this result would be very helpful.
The current world average reads \cite{HFAG}
\begin{equation}
Br( B^+ \to \tau^+ \nu_\tau)^{\rm HFAG \, 2013} = (1.14 \pm 0.22) \cdot 10^{-4} \; ,
\end{equation}
which is still larger than the theory prediction.
A detailed discussion of $B$-meson decays into final states with a $\tau$-lepton can 
be found in \cite{Soffer:2014kxa}.
\\       
The decay $B_s \to \mu^+ \mu^-$ proceeds in the standard model on the loop-level, either via penguins
or via a box diagram. Very recently the theory prediction was updated \cite{Bobeth:2013uxa}
including NNLO-QCD
corrections to obtain
\begin{eqnarray}
\bar{Br}(B_s \to \mu^+ \mu^-)^{\rm SM}  & = &  \left(3.65 \pm 0.23 \right) \cdot 10^{-9} \; ,
\\
\bar{Br}(B_d \to \mu^+ \mu^-)^{\rm SM}  & = &  \left(1.06 \pm 0.09 \right) \cdot 10^{-10} \; .
\end{eqnarray}
$\bar{Br}$ denotes the average time-integrated branching ratio that includes effects of
a finite value of $\Delta \Gamma_q$ \cite{DeBruyn:2012wk}.
The current experimental numbers read \cite{CMSandLHCbCollaborations:2013pla}
\begin{eqnarray}
\bar{Br}(B_s \to \mu^+ \mu^-)^{\rm Exp.}  & = &  \left(2.9 \pm 0.7 \right) \cdot 10^{-9} \; ,
\\
\bar{Br}(B_d \to \mu^+ \mu^-)^{\rm Exp.}  & = &  \left(3.6^{+1.6}_{-1.4} \right) \cdot 10^{-10} \; ,
\end{eqnarray}
which are averages from the CMS value \cite{Chatrchyan:2013bka}
and  
the LHCb value \cite{Aaij:2013aka}.
Standard model and experiment agree for the measured $B_s$-decay, but there is 
still room for substantial deviations, due to the large experimental uncertainties. 
The current bound on the $B_d$-decay is higher than the standard model expectation,
but here we have to wait for future more precise measurements
to see, whether there are some first hints of new physics in these decays or not.
Already the current experimental precision gives some interesting constraints
on 2HDM models or SUSY-models, in particular in the large $\tan \beta$-region.
\\          
In  \cite{Bobeth:2013uxa}  also the theory predictions of interesting and experimentally almost
unexplored decays like 
$B_q \to \tau^+ \tau^-$ were updated. 
 \begin{eqnarray}
\bar{Br}(B_s \to \tau^+ \tau^-)^{\rm SM} & = & \left(7.73 \pm 0.49 \right) \cdot 10^{-7} \; ,
\\
\bar{Br}(B_d \to \tau^+ \tau^-)^{\rm SM}  & = & \left(2.22 \pm 0.19 \right) \cdot 10^{-8} \; .
\end{eqnarray}
These decays are very  helpful for new physics searches,
see e.g. \cite{Bobeth:2014rda,Bobeth:2011st} and they are currently quite unconstrained. For
$B_s \to \tau^+ \tau^-$ no direct bound exists at all and for 
$B_d \to \tau^+ \tau^-$ there is a weak bound from BaBar \cite{Aubert:2005qw} (at $90\%$ C.L.)
\begin{eqnarray}
\bar{Br}(B_d \to \tau^+ \tau^-)^{\rm BaBar}  & < & 4.1 \cdot 10^{-3} \; .
\end{eqnarray}

  \subsection{Semi-leptonic decays}
Exclusive,  semi-leptonic $B$-meson decays are crucial for the determination of 
$ V_{cb}$ and $ V_{ub}$, which was already discussed in the Section 2.3.2.
For this purpose one investigates decays with electrons or muons in the final states.  
Having instead $\tau$-leptons, e.g. in 
$ B \to \bar{D}^{(*)} \tau^+ \nu_\tau$\footnote{Here we consider both  
$ B_d \to D^{-(*)} \tau^+ \nu_\tau$ and 
 $ B^+ \to \bar{D}^{0(*)} \tau^+ \nu_\tau$ and our final results are averages of the
two possibilities. For semi-leptonic decays the 
$B \to \bar{D}^*$-transition is roughly two times as common as the $B \to \bar{D}$ one.
}, one finds some 
deviations between experiment and theory. 
Usually the ratios
\begin{eqnarray} 
R(D^{(*)}) & = &  \frac{\Gamma (B \to \bar{D}^{(*)} \tau^+ \nu_{\tau})}
                       {\Gamma (B \to \bar{D}^{(*)} l^+ \nu_{l})}
\end{eqnarray}
are investigated, with $l$ denoting $e$ or $\mu$.
The theory prediction reads \cite{Fajfer:2012vx}
\begin{eqnarray}
R(D)^{\rm SM}& = & 0.296\pm 0.016 \; ,
\\
R(D^*)^{\rm SM} & = & 0.252\pm 0.003 \; .
\end{eqnarray}
BaBar measured in 2012 \cite{Lees:2012xj,Lees:2013uzd} the following values
\begin{eqnarray}
R(D)^{\rm BaBar}   & = & 0.440 \pm 0.058 \pm 0.042 \; ,
\\
R(D^*)^{\rm BaBar} & = & 0.332 \pm 0.024 \pm 0.018 \; ,
\end{eqnarray}
which differ sizably from the standard model expectation. Unfortunately
there exists no recent number from Belle. Updating the 2010 values from \cite{Bozek:2010xy}
the analysis in \cite{Lees:2013uzd} finds
\begin{eqnarray}
R(D)^{\rm Belle}   & = & 0.34 \pm 0.10 \pm 0.06 \; ,
\\
R(D^*)^{\rm Belle} & = & 0.43 \pm 0.06 \pm 0.06 \; ,
\end{eqnarray}
Here more data will be necessary to clarify this situation. 
A detailed discussion of the experimental situation can be found in \cite{Soffer:2014kxa}.
On the theory side, there was an ab-initio lattice calculation in \cite{Bailey:2012jg},
which obtained
\begin{eqnarray}
R(D)^{\rm Lattice}   & = & 0.316\pm 0.12 \pm 0.07 \; .
\end{eqnarray}
A similar result $R(D) = 0.31 \pm 0.02$ was obtained in \cite{Becirevic:2012jf}, being still lower than 
the experimental number.
\\
There is a second class of semi-leptonic decays that triggered a lot of interest:
$ B \to  K^{(*)} \mu^+ \mu^-$. In contrast to $B \to \bar{D}^{(*)} \tau^+ \nu_\tau$, which is 
a tree-level decay in the standard model, the decay $ B \to  K^{(*)} \mu^+ \mu^-$
is triggered by a $ b \to s \mu^+ \mu^-$-penguin or box diagram, as the decay $B_s \to \mu^+ \mu^-$.
\\
Using $1 fb^{-1}$ of data, LHCb measured in 2005 \cite{Aaij:2012cq}
a 4.4 $\sigma$ deviation of the isospin asymmetry $A_I$, defined as
\begin{equation}
A_I = \frac{Br(B_d \to K^0 \mu^+ \mu^-) - \frac{\tau (B_d)}{\tau (B^+)}Br(B^+ \to K^+ \mu^+ \mu^-)}
{Br(B_d \to K^0 \mu^+ \mu^-) + \frac{\tau (B_d)}{\tau (B^+)}Br(B^+ \to K^+ \mu^+ \mu^-)} \; ,
\end{equation}          
from the tiny standard model prediction \cite{Feldmann:2002iw,Khodjamirian:2012rm,Lyon:2013gba}. 
In 2014 this measurement was updated \cite{Aaij:2014pli} with the full data set of  
$3 fb^{-1}$ and the deviation disappeared (it was reduced to 1.5 standard deviations).
On the other hand, the measured branching fractions of the four $B \to K^{(*)}\mu^+ \mu^- $ decays
\cite{Aaij:2014pli,Aaij:2013iag} and the decay $B_s \to \phi \mu^+ \mu^- $ \cite{Aaij:2013aln}
have all lower values than the standard model expectations \cite{Horgan:2013hoa,Horgan:2013pva}.
\\
The same large data set was used in \cite{Aaij:2014tfa}  to  perform an angular analysis of 
charged and neutral $B \to K \mu^+\mu^-$ decays and no deviation from the small
standard model expectations
\cite{Grinstein:2004vb,Bobeth:2007dw,Egede:2008uy,Khodjamirian:2010vf,Bobeth:2011nj} was found.
\\
The angular analysis of the decay $B \to K^{0*} \mu^+\mu^-$ was published in 2013
\cite{Aaij:2013qta} with a data set of $1 fb^{-1}$. This decay can be expressed in terms of the 
eight form factor like
parameters $F_L$, $S_{3-9}$ \cite{Altmannshofer:2008dz}. In \cite{Descotes-Genon:2013vna}
it was suggested to use instead the parameters $P_j' = S_j/\sqrt{F_L(1-F_L)}$ for $j = 4,5,6,8$,
because some hadronic contributions cancel. LHCb measured \cite{Aaij:2013qta}
the four parameters $P_j'$ in six
different $q^2$-bins and from the 24 measurements 23 agreed with the standard model, while the
24th one, related to $P_5'$, deviated by 3.7 $\sigma$. This discrepancy triggered a lot of theoretical interest, 
see e.g. \cite{Descotes-Genon:2013wba,Altmannshofer:2013foa,Gauld:2013qba,Hambrock:2013zya,Buras:2013qja,Gauld:2013qja,Beaujean:2013soa,Hurth:2013ssa,Matias:2014jua}. Future investigations
of the hadronic uncertainties as well as the results of the 
$3 fb^{-1}$ data set will give further clues.

  \subsection{Non-leptonic decays}
Hadronic $B \to D K$-decays  can be used to extract the CKM angle $\gamma$ directly, see e.g.
the GLW-method \cite{Gronau:1990ra,Gronau:1991dp},
the ADS-method \cite{Atwood:1996ci,Atwood:2000ck}
or the GGSZ-method \cite{Giri:2003ty}   .
These methods provide a clean consistency check of the CKM picture with decays that 
proceed only via tree-level and are thus expected to be less sensitive to new 
physics effects.
Currently values from LHCb \cite{Aaij:2013zfa}, BaBar\cite{Lees:2013nha} and Belle
\cite{Trabelsi:2013uj}
are available 
\begin{eqnarray}
\gamma^{\rm LHCb } & = & \left(72.0 ^{+14.7}_{-15.6}\right)^\circ \; ,
\\
\gamma^{\rm BaBar} & = & \left(69 ^{+17}_{-16}\right)^\circ \; ,
\\
\gamma^{\rm Belle} & = & \left(68 ^{+15}_{-14}\right)^\circ \; ,
\end{eqnarray}
which can be compared with the CKM-fit result \cite{CKMfitter, UTfit}
\begin{eqnarray}
\gamma^{\rm CKMfitter} & = & \left(69.7 ^{+1.3}_{-2.8}\right)^\circ \; .
\end{eqnarray}
These numbers agree, but the precision of the direct determination is not yet comparable to
the indirect one. Here it will be very interesting to see what happens, 
when the experimental precision
is improving.
\\
The angle $\beta$ can be obtained by studying the decay
$B_d \to J / \psi K_S$. In contrast to the previous case, which was dominated by 
tree-level contributions, the dependence on the CKM-angle $\beta$ arises from the
interference between mixing and decay and is thus related to a loop process.
The values for $\beta$ from LHCb \cite{Aaij:2012ke}, BaBar \cite{Aubert:2009aw}
and Belle \cite{Adachi:2012et}
read
\begin{eqnarray}
\sin 2\beta^{\rm LHCb } & = & 0.73  \pm 0.07  \pm 0.04 \; ,
\\
\sin 2\beta^{\rm BaBar} & = & 0.687 \pm 0.028 \pm 0.012 \; ,
\\
\sin 2\beta^{\rm Belle} & = & 0.667 \pm 0.023 \pm 0.012 \; ,
\end{eqnarray}
which can again be compared with the CKM-fit result \cite{CKMfitter, UTfit}
\begin{eqnarray}
\sin 2\beta^{\rm CKMfitter} & = & 0.775^{+0.020}_{-0.049}\; .
\end{eqnarray}
This deviation caused some discussion in the literature, see e.g. \cite{Lenz:2010gu}
and references therein. It might be related to new physics in $B_d$-mixing and/or to the extraction
of $V_{ub}$.
\\
The related decays $B_s \to J / \psi K^+ K^-, J / \psi \pi^+ \pi^-,...$ can be used to 
extract the mixing phase $\beta_s$ in the $B_s$-system, which is predicted to be 
very small in the standard model
\cite{CKMfitter,UTfit}.
\begin{equation}
\beta_s^{\rm CKMfitter} = 0.01821^{+0.00081}_{ -0.00079} \;.
\end{equation}
Using $1fb^{-1}$ of data LHCb found \cite{Aaij:2013oba}
\begin{equation}
\beta_s^{\rm LHCb} = 0.01 \pm 0.07 \pm 0.01 \;.
\end{equation}
Both numbers are consistent, but the experimental uncertainty is still considerably 
larger than the theoretical one. The phase $\beta_s$ should not be mixed up, with the
mixing phase $\phi_s$ defined below Eq.(\ref{deltagamma}), see e.g. the ``note added'' in
\cite{Lenz:2007nk,Lenz:2008xt}.
\\
Finally we would like to briefly discuss direct CP violation in hadronic $D$-meson decays - see
\cite{Lenz:2013pwa} and references therein for a more detailed discussion.
$\Delta A_{CP}$  is defined as the difference of the CP asymmetries of a neutral $D$-meson decaying
into $KK$ and 
$\pi \pi$ final states.
\begin{equation}
\Delta A_{CP} := A_{CP} (D^0 \to K^+K^-) - A_{CP}(D^0 \to \pi^+\pi^-) 
\; .
\end{equation}
The first measurements in 2011 
\cite{Aaij:2011in,Collaboration:2012qw,Ko:2012px} gave a combined value
of 
\begin{equation}
\Delta A_{CP} =  - 0.678 \pm 0.147 \% \, .
\label{DeltaACPold}
\end{equation}
Such a large value was quite unexpected in the standard model. LHCb performed, however, subsequent
measurements where the significance went down \cite{LHCupdate} and also some which resulted in a 
different sign \cite{Aaij:2013bra}. Taking these new numbers into
account the new combination turns out to be \cite{HFAG}
      \begin{equation}
      \Delta A_{CP} =  - 0.329 \pm 0.121 \% \, .
      \label{DeltaACPnew}
      \end{equation}
      The statistical significance for CP violation in hadronic $D$ decays
      went now down considerably, but the 
      central value is still larger than to be expected naively in the 
      standard model. Here clearly  further experimental input is 
      needed to settle this issue.

 \begin{figure}[h]
  \centering
  \fbox{
    \includegraphics[width=0.6\textwidth]{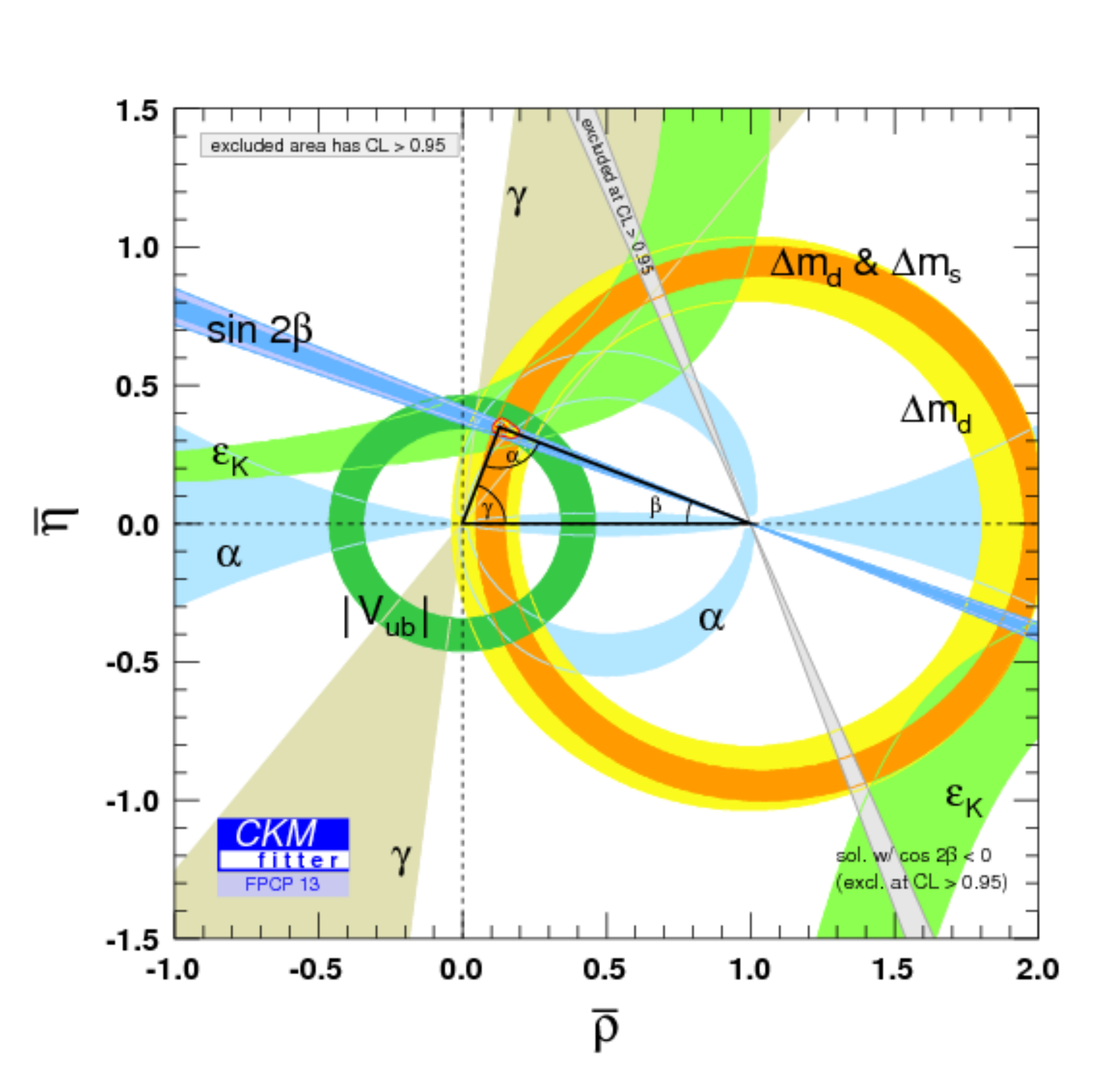}
  }
  \caption{The current status of the CKM fit taken from \cite{CKMfitter}, similar 
results can be obtained from \cite{UTfit}.}
\label{fig:ckmfit}
\end{figure}

\section{Consequences for new physics models}
\label{newphysics}

A general test of the consistency of the CKM picture is provided by the usual fit 
of the unitarity triangle, see e.g. \cite{CKMfitter} and \cite{UTfit}. Here observables like
$V_{ub}$, $\Delta M_d$, $\Delta M_s$, $\sin 2 \beta$ and CP-violation in the Kaon system,
$\epsilon_K$, are included. 
As can be seen from Fig.\ref{fig:ckmfit} the currently available amount of flavour
data is very well compatible with the CKM paradigm. Nevertheless, this does not exclude
the possibility of having sizable new physics contributions in the flavour sector, which will be
investigated below.

     \subsection{Model independent search for new physics}
\begin{figure}[h]
  \centering
  \fbox{
    \includegraphics[width=0.45\textwidth]{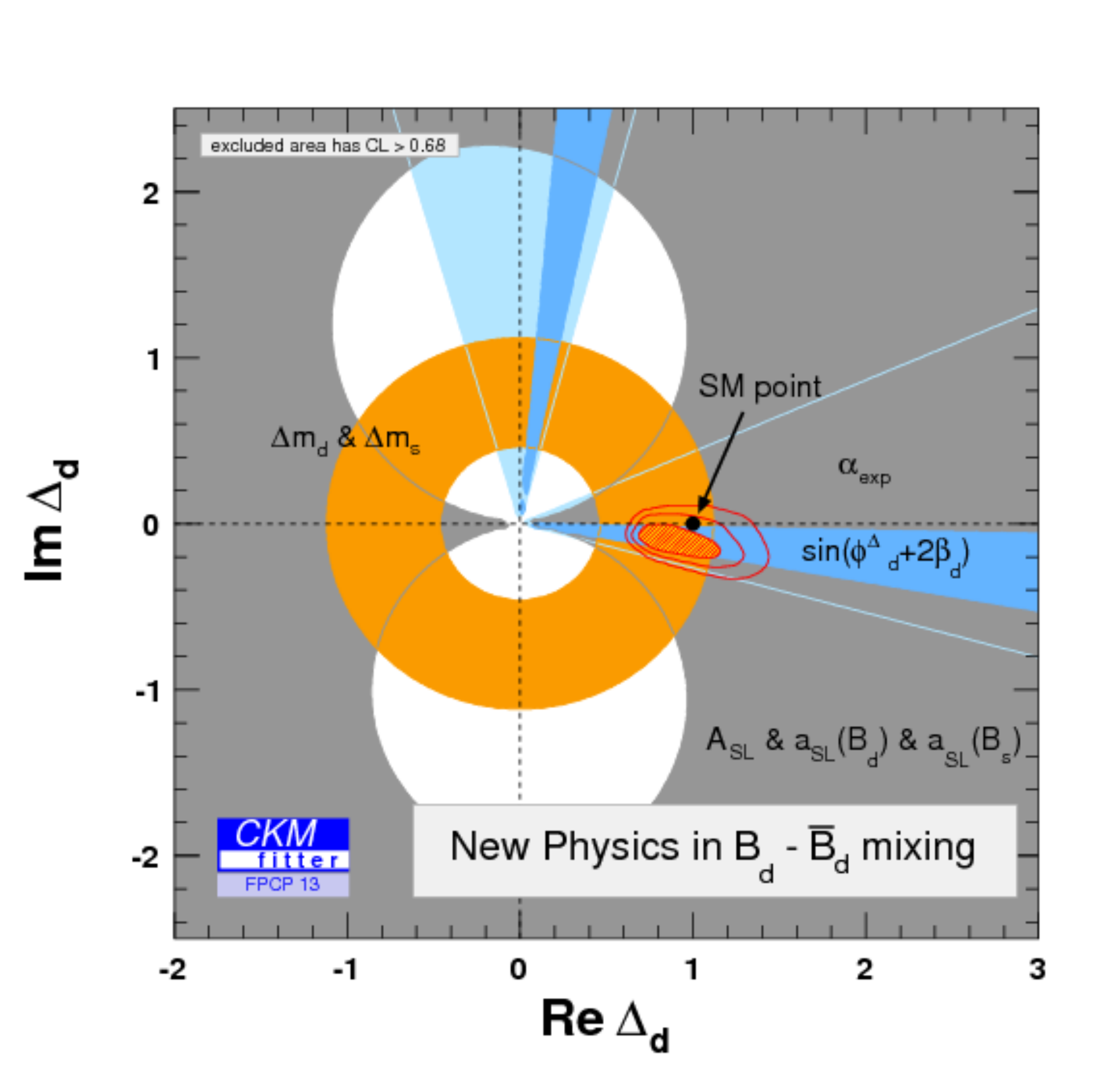}
    \hfill
    \includegraphics[width=0.45\textwidth]{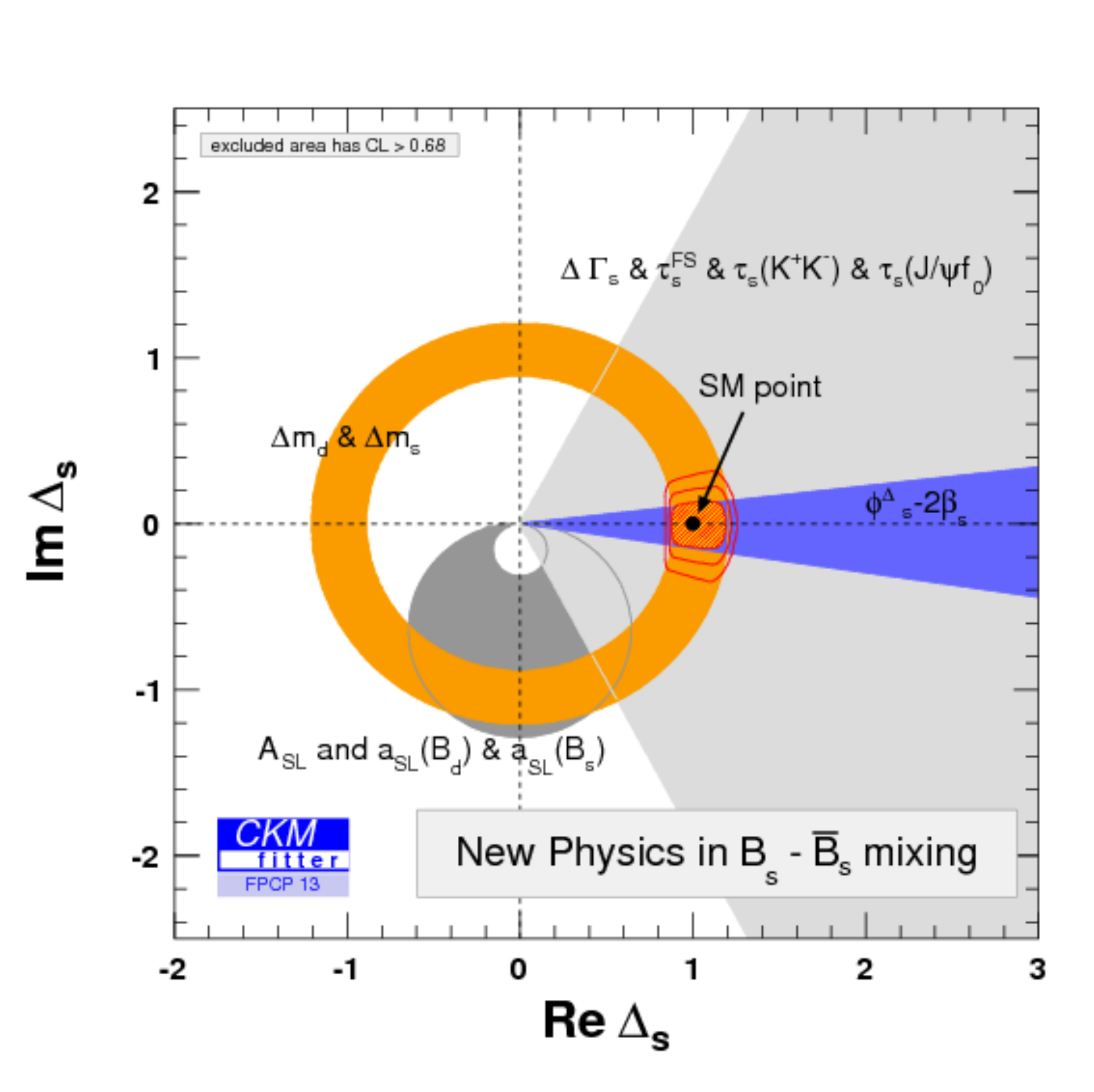}
    }
    \caption{Allowed space for new physics effects in $B_d$- and $B_s$-mixing. The figures are taken 
             from \cite{CKMfitter}, they are an update of \cite{Lenz:2012az}.}
    \label{fig:NPB}
\end{figure}

\begin{figure}[h]
  \centering
  \fbox{
    \includegraphics[width=0.5\textwidth]{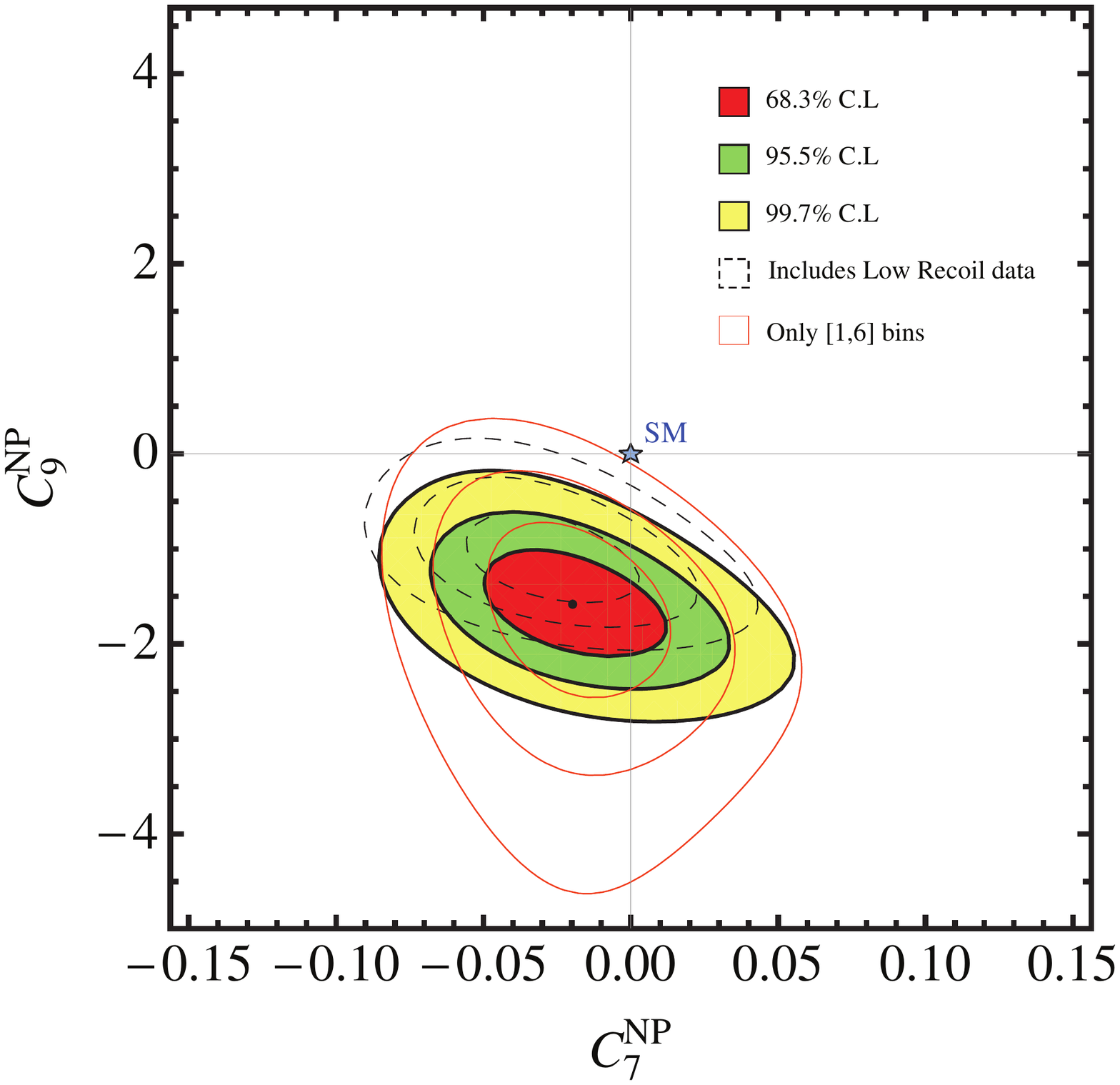}
    }
    \caption{Constraints on the new physics contributions to the Wilson coefficients $C_7$ and $C_9$ 
             from penguin induced $b \to s$ transitions. The figure is taken from 
             \cite{Descotes-Genon:2013wba}.}
    \label{fig:P5p}
\end{figure}

\begin{figure}[h]
  \centering
  \fbox{
    \includegraphics[width=0.45\textwidth]{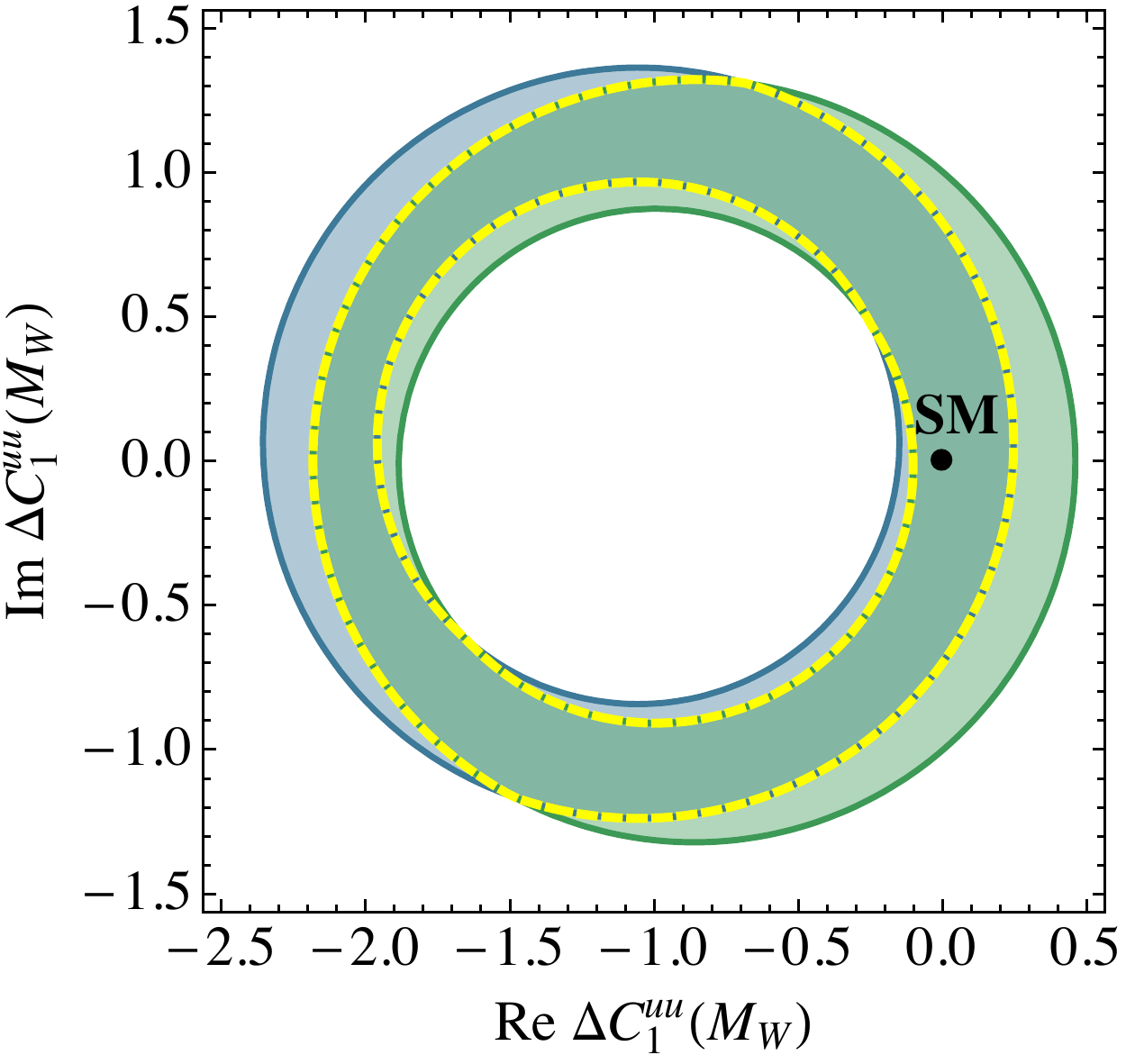}
    \hfill
    \includegraphics[width=0.45\textwidth]{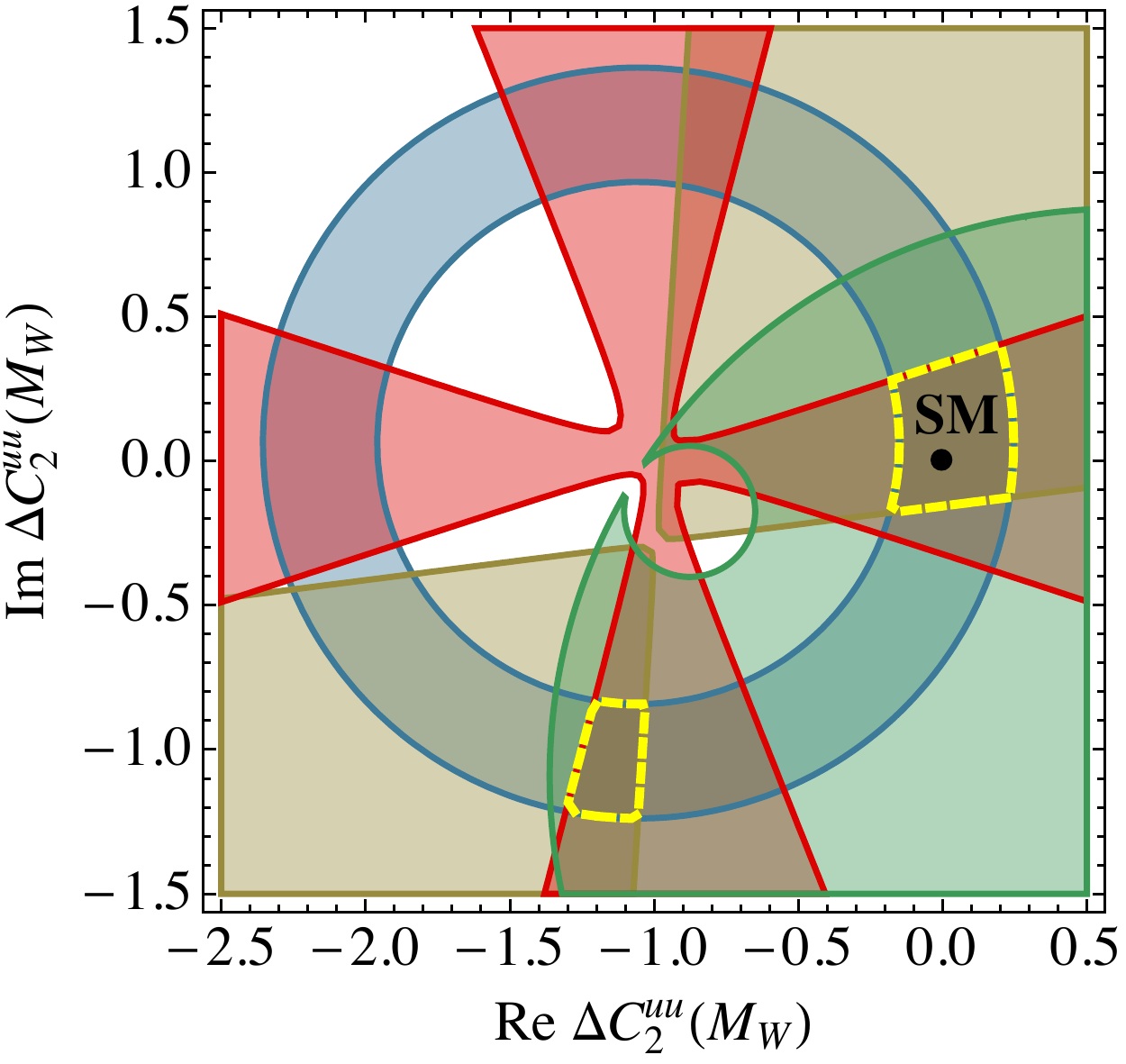}
    }
    \caption{Allowed space for new physics effects in the Wilson coefficients $C_1$ and $C_2$ of 
             the current-current operators for the case of  $b \to u \bar{u} d$-transitions.
             The figures are taken 
             from \cite{Bobeth:2014rda}.}
    \label{fig:NPC12}
\end{figure}

There are different ways of performing model independent searches for new physics effects.
Mixing seems to be a promising place to search for beyond standard model effects, because it
is a loop effect. In \cite{Lenz:2010gu} and \cite{Lenz:2012az} new physics effects in 
mixing were estimated under the assumption of having only considerable effects in mixing,
in $M_{12}$, 
while the tree-level decay amplitudes are dominated
by standard model contributions, i.e. the relation between the {\it true} values
of $M_{12}$ and $\Gamma_{12}$ and their standard model counterparts  
$M_{12}^{\rm SM}$ and $\Gamma_{12}^{\rm SM}$ reads
\begin{eqnarray}
M_{12}^q & = & M_{12}^{q, \rm SM} \Delta_q \; ,
\\
\Gamma_{12}^q & = & \Gamma_{12}^{q \rm SM} \; ,
\end{eqnarray}
where $\Delta$ is an arbitrary complex number, encoding the new physics contribution. 
This assumption corresponds also to e.g. neglecting new penguin contribution in the decays
$B_d \to J / \psi K_S$ and $B_s \to J / \psi \phi$ and thus the values for $\beta$ and $\beta_s$
give information on the mixing phase $\phi_q = \arg (- M_{12}^q/\Gamma_{12}^q) = 
\phi_q^{\rm SM} + \phi_q^\Delta$.
Combing all available data till 2013 for
the $B_d$-system one gets the bounds shown on the left hand side of Fig.\ref{fig:NPB}, 
while the bounds on the $B_s$-system are displayed on the right hand side of  Fig.\ref{fig:NPB}.
In the $B_d$-system the fit prefers a negative value of the phase of $\Delta_d$, but the deviation
from the standard model is less than 2 $\sigma$. On the other hand 
values of $\phi_d^\Delta$ of about $-10^\circ$, which would be a quite large new physics 
contributions, are clearly not ruled out. In the $B_s$-system the fit prefers the standard model
value, but it leaves also space for considerable deviations. In that case 
up to $\pm 20^\circ$ are not rule out yet.
\\
Similar studies were performed for the Wilson coefficients  $C_7$, $C_9$ and $C_{10}$ 
of $b \to s$-penguins transitions.
In \cite{Descotes-Genon:2013wba} $P_5'$ could be explained by a negative shift in $C_9$, while 
$C_7$ stays standard model like, see e.g. Fig.\ref{fig:P5p}.
Similar, but sometimes less pronounced results - i.e. better consistent with the standard model - 
were found e.g. in \cite{Altmannshofer:2013foa,Hambrock:2013zya,Beaujean:2013soa,Hurth:2013ssa}.
\\
A complementary study was performed in \cite{Bobeth:2014rda}. 
Here the space of new physics effects in the Wilson coefficients $C_1$ and $C_2$ of 
tree-level decays like $b \to c \bar{u} d$, $b \to c \bar{c} d$, $b \to u \bar{c} d$ and 
$b \to u \bar{u} d$ was investigated and it was found that notable deviations from the standard model
expectations are still possible, see Fig.\ref{fig:NPC12} for the bounds on $C_1$ and $C_2$ for
the case of $b \to u \bar{u} d$-transitions.

    \subsection{Decay channel  independent search for new physics}
An even more general BSM search strategy might be provided by the study of inclusive non-leptonic
decays in the spirit of the missing charm puzzle \cite{Lenz:2000kv}.
This was advocated again in \cite{Krinner:2013cja}.
Comparing experiment and theory for partially summed branching ratios, 
like $Br(b \to {\rm no \; charm})$ one could get information on all invisible decay modes, like e.g.
$B \to \tau \tau$ or even more fancy possibilities like an invisible decay into light dark matter
particles. The latest experimental studies in that direction, a determination of $n_c$ - the 
average number of charm quarks per $b$-decay - date back to 2006 \cite{Aubert:2006mp}. Non-leptonic
inclusive decays might also gain some insight into CP violation, see e.g. \cite{Lenz:1998qp}.

     \subsection{Model dependent search for new physics}
There exists an enormous amount of literature about model dependent searches 
for new physics effects in flavour observables. A corresponding discussion
is beyond the scope of this review.
 
\section{Conclusion}   
\label{conclusion}
\setcounter{equation}{0} 
We have discussed a selected choice of topics in flavour physics in order to give
an idea of the current status of the field. For different choices of topics see e.g.
the reviews
\cite{Gersabeck:2012xk,Gersabeck:2012rp,Borissov:2013yha,Gershon:2013aca,Nir:2013maa,Isidori:2014rba}.
Splitting up the current investigations in three areas: 
testing the theoretical tools, determining standard model parameters
and search for new physics we come to the following conclusions:
\begin{enumerate}
\item  Testing of our theoretical tools: Applying the HQE for lifetimes of $b$- and $c$-hadrons as well
       as for $\Delta \Gamma_s$ gives very promising results and there is no huge space for
       violations of duality anymore. Old discrepancies like the 
       $\Lambda_b$-lifetime have finally been settled experimentally, unknown quantities
       like $\Delta \Gamma_s$ have been measured for the first time in perfect agreement with
       theory and it looks like even $D$-meson lifetimes might be described by the HQE.
       Since it is clear now that the HQE works, the new question is, how precise is the HQE. 
       To answer that lattice results for many of the arising observables are urgently needed.
       \\
       There are, however, also some areas where it is not clear yet, how to describe them
       in theory. For $D$-mixing it might be worthwhile to push the HQE to its limits and determine
       dimension nine and dimension twelve contributions. For more complicated problem
       like $\Delta A_{CP}$ it is almost unclear, how to proceed, although there are
       some interesting ideas related to lattice, see e.g. \cite{Hansen:2012tf}
\item Determining standard model parameters: in that field a huge progress was made, as can be seen
      e.g. in Fig.\ref{fig:ckmfit} or in the current precision of the CKM-element $V_{cb}$. Among many
      other observables, future 
      measurements of $\gamma$ will provide a clean cross-check of the CKM-picture.
\item Search for new physics: here we have three results: a) most observables are standard
      model like, b) there is nevertheless still a lot of space for effects beyond the standard model
      and c) there is still a notable number of remaining discrepancies.
      \begin{enumerate}
      \item Most observables are standard model like:
            \\
            This should actually not be a source of disappointment, since it is an amazing
            success of our theory. Complicated loop observables like $\Delta M_q$, $\Delta \Gamma_s$
            or $ b \to s \gamma$ are described very well by the  standard model. Even
            very rare decays like $B_s \to \mu^+   \mu^-$ have been predicted many years before
            their discovery (e.g. Buras quoted 1998 \cite{Buras:1998raa} 
            in his Les Houches Lectures a value
            of $(3.4 \pm 1.2) \cdot 10^{-9})$. This is a real impressive success of the standard model. 
      \item Despite being standard model like, there is still a lot of room for new physics in 
             observables like
             $B_s \to \mu^+  \mu^-$, $Br_{sl}$, $\Delta M_q$, $\Delta \Gamma_d$ , $a_{sl}^q$,
             $\beta_s$ and $\beta$. Here more precise measurements as well as more precise theoretical
             studies will either shrink further the space for new physics effects or find 
             the first convincing hints.
      \item There are still remaining discrepancies, e.g.
            $V_{ub}$, the dimuon asymmetry, $Br(B^+ \to \tau^+ \nu_\tau)$, $R(D^{(*)})$, $Br(B_d \to 
            \mu^+ \mu^-)$,
            $Br(B \to K^{(*)} \mu^+ \mu^-)$, $Br(B_s \to \phi \mu^+ \mu^-)$, $P_5'$, 
            $\Delta A_{CP}$, $\beta$,...
             Here again more precise measurements as well as more precise theoretical
             studies will shed light on the origin of this discrepancies.
      \end{enumerate}
      For future searches we also would like to stress some lesser known, but nevertheless promising
      observables like $B \to \tau^+ \tau^-$,  $\Delta \Gamma_d$ and inclusive non-leptonic decays.
\end{enumerate}
So we are heading towards exciting, but strenuous times in flavour physics.

\section*{Acknowledgements}   
    
I would like to thank Gilberto Tetlalmatzi-Xolocotzi for proofreading the manuscript and Alan Martin
for suggesting this review.


\end{document}